\documentclass[prb,epsfig,aps,amsmath,twocolumn]{revtex4}

\usepackage{graphicx}% Include figure files
\usepackage{dcolumn}% Align table columns on decimal point
\usepackage{bm}% bold math

\begin{document}

\title{Model study of adsorbed metallic quantum dots: Na on Cu(111)}

\author{T.~Torsti$^{1,2}$, V.~Lindberg$^3$, M.~J.~Puska$^1$, and
B.~Hellsing$^4$}

\affiliation{$^1$Laboratory of Physics, Helsinki University of Technology, P.O. Box 1100, FIN-02015 HUT, Finland}
\affiliation{$^2$CSC - Scientific Computing Ltd., P.O. Box 405, FIN-02101 Espoo, Finland}
\affiliation{$^3$Department of Physics, V\"axj\"o University, SE-35195, V\"axj\"o, Sweden}
\affiliation{$^4$Department of Experimental Physics,Chalmers and G\"oteborg University, SE-412 96, G\"oteborg, Sweden}

\date{\today} 

\begin{abstract}
We model electronic properties of the second monolayer Na adatom 
islands (quantum dots) on the Cu(111) surface covered homogeneously 
by the first Na monolayer. An axially-symmetric three-dimensional 
jellium model, taking into account the effects due to
the first Na monolayer and the Cu substrate, has been developed. The
electronic structure is solved within the local-density approximation 
of the density-functional theory using a real-space multigrid method. 
The model enables the study of systems consisting of thousands 
of Na-atoms. The results for the local density of states are compared 
with differential conductance ($dI/dV$) spectra and constant current 
topographs from Scanning Tunneling Microscopy. 
\end{abstract}

%-------------------------------------------------------------------
%
%

\maketitle

%\newpage

\section{Introduction} 

At certain faces of metals, such as the (111) face of noble metals, the
surface electron  states are confined to the vicinity of the top layer by the
vacuum barrier on the vacuum side and the bandgap on the substrate side
\cite{Luth}. The electrons in these surface states form a two-dimensional nearly free
electron gas \cite{Gartland,Heimann,Zangwill,Ashcroft-Mermin}. 
It has also been observed that when adsorbing one to several monolayers of
alkali atoms on these surfaces a manifold of discrete standing wave
states, so called quantum well states (QWS), perpendicular to the
surface are formed. \cite{Lindgren-Wallden80,Lindgren-Wallden87}. These states can be
detected, for instance, in photoemission spectroscopy (PES) \cite{Carlsson97},
inverse photoelectron spectroscopy (IPES) \cite{Dudde}, two photon
photoemission spectroscopy (2PPES) \cite{Fisher} and scanning tunneling
microscopy (STM) \cite{Kliewer00}. A large amount of experimental data is
available for the system Na on Cu(111) 
\cite{Lindgren-Wallden80,Lindgren-Wallden87,Carlsson97,Dudde,Fisher,Kliewer00}. The electronic
structure and dynamics for this system have also been investigated by 
first-principles theoretical calculations \cite{Carlsson,Hellsing2000}.

These localized surface states are of great interest since they play an important role
in many physical processes like epitaxial growth \cite{Memmel}, surface
catalysis \cite{Bertel,Lauritsen}, molecular ordering \cite{Stranick} and
adsorption \cite{Bertel97}. Experimental tools like STM and PES play an
important role in the investigations, since they enable spatial and
spectroscopic resolution of the electron states.

%One important discovery is the confinement of surface state electrons in
%nanoscale structures, so called quantum corrals, by deliberate assembled
%adatoms to enclosed structures by STM \cite{Crommie}. 
One important discovery is the confinement of surface state electrons in
so called quantum corrals. These man-made nanoscale structures are formed  
by deliberately assembling adatoms to enclosed structures by STM \cite{Crommie}. 
 Due to the small
size of the corrals, quantum effects are present, and both spatial and
spectroscopic properties of the confined states can be studied
experimentally. A natural way of forming low-dimensional structures on metal
surfaces is by controlled growth of epitaxial layers. With an appropriate
choice of deposition and annealing temperatures small islands, so called
quantum dots (QD), with variable shapes and sizes may form \cite{Roder}. The
advantage of these structures, in comparison with the corrals, is that they are
relatively stable at low temperatures. This enables the imaging and
investigation of their properties without inducing structural damage. One quantum
mechanical effect of the confinement is the increase of the surface state energy
which in turn may lead to the depopulation of the surface
state band and thereby changes in the surface properties.

In this paper we present calculated results for the electronic structure
of Na on Cu(111), with the emphasis on describing the real-space
resolved density of states nearby a sodium QD adsorbed on a
sodium-covered Cu(111) surface. Previously, an all-electron density-functional 
theory (DFT) study of a free-standing Na layer in vacuum has been 
presented \cite{Wimmer}. More recently, a DFT calculation using ultrasoft 
pseudopotentials for the free-standing Na layer as well as for the layers 
adsorbed on Cu(111) 
have been presented\cite{Carlsson}. A simple free electron model calculation for 
a free-standing Na QD has already been published by two of the present
authors\cite{vanja1}.
 
The calculations in this work are made in the context of the DFT \cite{lang83,dft}.
More specifically, the Rayleigh-quotient multigrid (RQMG) method in
axial symmetry\cite{mgarticle1,mika} is used for the numerical solution of the ensuing
Kohn-Sham equations.
%MIKA (Multigrid Instead of the K-spAce)
%\cite{mgarticle1,mika} real-space multigrid package. 
The electron-ion interaction
is simplified using the jellium model \cite{lang83}, where the ions are
replaced by a rigid positive background charge of constant density. 
This model has provided basic physical understanding of the electronic structures of
simple metal  surfaces \cite{lang83} thin metal films \cite{Schulte}, vacancies and voids inside metals
\cite{manninen75} and finite clusters of simple metal atoms \cite{heerbrack}.
Recently, also uniform cylindrical nanowires have been studied within the jellium 
model \cite{zabala1,zabala2,zabala3}.

The paper is organized as follows: Section \ref{naoncu} gives a short review over 
experimental results for the system Na on Cu(111), Section \ref{compmet} 
describes  the computational method used in the calculations, Section \ref{model} 
discusses the details of our jellium model and in Section \ref{meas} the results 
are presented and comparisons are made  with experimental findings. Finally, 
Section \ref{conclusion} gives the conclusions.

\section{Na on Cu(111)} \label{naoncu}

Alkali metals adsorbed on the closed-packed (111) surface of metals form
hexagonal structures at saturated monolayer coverages, following approximately
the underlying substrate structure \cite{Diehl}. The first monolayer of Na on
Cu(111) is observed to saturate at the coverage of $\Theta = 4/9 \approx 0.44$
\cite{Tang}, corresponding to 4 Na atoms per 9 surface Cu atoms. The Na atoms thus
form a hexagonal (3/2 $\times$ 3/2) structure and the Na atom spacing of $7.43 \;\mathrm{a_0}$
is comparable to the atomic distance of $6.92 \;\mathrm{a_0}$ in bulk Na.

The adsorption of Na atoms on the Cu(111) surface will induce a charge
redistribution at the interface between the adlayer and the substrate. It has
been seen from photoemission experiments \cite{Fisher,Lindgren-Wallden88}
that when the Na coverage is increased the Cu Shockley surface state decreases in
energy. For coverages above $\Theta \approx 0.11$ the surface state is
shifted below the lower band edge of the local band gap of the Cu(111) surface, and
is no  longer visible in photoemission experiments. Two-photon photoemission
experiments \cite{Fisher} indicate other unoccupied Na-induced states in the
local bandgap at the Cu(111) surface, which will also decrease in energy with
increasing Na coverage. For higher coverages, the lowest of these states will
be down-shifted below the Fermi energy, and thus get occupied. At the saturated
monolayer coverage, this state will be located about $0.1\;\mathrm{eV}$ 
\cite{Fisher,Carlsson,Lindgren-Wallden88} below the Fermi energy at the $\overline{\Gamma
}$-point of the surface Brillouin zone (SBZ). The corresponding next lowest
state will be located $2.1\;\mathrm{eV}$ above the Fermi level \cite{Fisher} at the
$\overline{\Gamma }$-point. The lower of the two states has one node in the
$z$-direction, while the higher has two.

If the Na atom deposition continues after the first monolayer (ML) is completed, a second
layer will start to grow. Recent STM measurements 
\cite{Kliewer_Berndt_SS01,Kliewer_Berndt_PRB01} indicate that the second monolayer of Na grows via the
formation of compact islands with hexagonal atomic arrangement.  Normal
photoemission experiments \cite{Carlsson97,Carlsson94} indicate that when
the second monolayer grows the emission intensity due to one-monolayer states
decreases gradually, and for coverages above 1.3 ML a new peak, approximately
$0.1\;\mathrm{eV}$ above the Fermi energy appears. This peak is ascribed to the
two-monolayer thick parts, and the energy is shifted to somewhat lower values
as the coverage is increased.

Some theoretical attention has also been paid to Na on Cu(111),
including the island growth. Free-electron model calculations have been performed for
circular  Na \cite{vanja1} as well as hexagonal Ag \cite{CrampinPRL,CrampinSS}
 and Na \cite{Berndt} free-standing islands. All-electron  calculations for
an unsupported monolayer of Na \cite{Wimmer} and first-principles slab
calculations for one atomic Na layer in (2 $\times$ 2) and  (3/2 $\times$ 3/2)
adsorbate structures on Cu(111)  \cite{Carlsson} have been presented. 
%An early jellium study by Schulte \cite{Schulte} on thin metal films should also 
%be mentioned here.
In the present paper, we report jellium model calculations for 
an unsupported monolayer of Na and a cylinder shaped free-standing Na QD. We also present
two-density-jellium calculations for the system Na on Cu(111), where we have
modeled the underlying Cu(111) substrate by using a lower density slab to
mimic the decay of the surface states into the substrate. Comparison is made
with experiments and previous theoretical calculations.

\section{Computational methods}  \label{compmet}

In the Kohn-Sham scheme of DFT, one solves the electron density $n({\bf r})$ of the system
selfconsistently from a set of equations. One of these equations is the single particle 
Schr\"odinger equation. The models used in this work are axially symmetric. Thus 
the Schr\"odinger equation is separable, and the wave functions can be written as products
\begin{equation}
\psi_{m{\bf k}n}(r,z,\phi) = e^{im\phi}U_{m{\bf k}n}(r,z).
\end{equation}
Above, $m$ is the azimuthal quantum number implied by the axial symmetry while $n$ 
differentiates between orthogonal states with same $m$ and ${\bf k}$.
In the calculations involving the infinite monolayer, two {\bf k}-vectors are 
used as explained in Sec. \ref{model}. 
The external potential of the systems studied in this work is caused by the positive background 
charge $n_+({\bf r})$.
% and the term $V_{stab}$ of the stabilized-jellium model
%\cite{stab}. $V_{stab}$ is a finite constant inside the volume of the background charge
%and vanishes beyond that volume. $V_{stab}$ stabilizes the jellium at the given bulk density.
The effective potential $V_{\rm eff}$ includes also the Hartree potential of the electron
density and the  exchange-correlation potential $V_{\rm XC}$, which we treat in the
local density approximation (LDA) \cite{xc}. The electron density $n({\bf r})$ is obtained by
summing single-electron densities with the occupation numbers $f_{m{\bf k}n}$. 
The degeneracies of the states are taken into account by the factor $(2-\delta _{0m})$ and the occupation
numbers $f_{m{\bf k}n}$  obey the Fermi-Dirac statistics with a Fermi level 
($E_F$) so that the system is neutral. 
A finite  temperature of 1200 K is used to stabilize the solution of the
set of equations. Thus, in the present axial symmetry (${\bf r} = (r,z)$)
the Kohn-Sham equations read as
\begin{widetext}
% The splitted eq
%\begin{equation}
%\begin{split}
%\label{kohnshameq}
%% \left(-\frac{1}{2}\nabla^2 +
%%        V_{\rm eff}({\bf r})
%%       \right) \Psi_i = \epsilon_i \Psi_i,
%-\frac{1}{2}  \left(\frac{1}{r}\frac{\partial}{\partial r} + 
%\frac{\partial^2}{\partial r^2}  - \frac{m^2}{r^2}  \frac{\partial^2}{\partial z^2} + 2V_{eff} \right) &  \\
% \times U_{m{\bf k}n}({\bf r})  =
%  \varepsilon_{m{\bf k}n}  U_{m{\bf k}n} & ({\bf r})
% \\
%\end{split}
%\end{equation}
% The widetext version
\begin{equation}
\label{kohnshameq}
% \left(-\frac{1}{2}\nabla^2 +
%        V_{\rm eff}({\bf r})
%       \right) \Psi_i = \epsilon_i \Psi_i,
-\frac{1}{2}  \left(\frac{1}{r}\frac{\partial}{\partial r} + 
\frac{\partial^2}{\partial r^2}  - \frac{m^2}{r^2}  \frac{\partial^2}{\partial z^2} + 2V_{eff} \right) 
  U_{m{\bf k}n}({\bf r})  =
  \varepsilon_{m{\bf k}n}  U_{m{\bf k}n} ({\bf r})
\end{equation}
\begin{equation}
\label{densitydef}
%  n({\bf r}) = \sum\limits_i^N f_{m{\bf i}}|U_{m{\bf i}}({\bf r})|^2,
  n({\bf r}) = 2\sum_{m{\bf k}n} (2-\delta_{0m}) f_{m{\bf k}n}|U_{m{\bf k}n}({\bf r})|^2,
\end{equation}
\begin{equation}
\label{veffdef}
V_{\rm eff}({\bf r}) = \Phi({\bf r}) 
 + V_{\rm XC}({\bf r}), % + V_{stab},
\end{equation}
\begin{equation}
%\begin{split}
\left(\frac{1}{r}\frac{\partial}{\partial r} + \frac{\partial^2}{\partial r^2} +
 \frac{\partial^2}{\partial z^2}\right)  \Phi = 
 -4\pi  \left[n_-({\bf r}) - n_+({\bf r})\right]. 
\label{poisson}
%  V_{\rm H}({\bf r}) = \int \frac{n({\bf r}')}
%         {|{\bf r}-{\bf r}'|}d{\bf r}',
%\end{split}
\end{equation}
%\begin{equation}
%\label{vxcdef}
%  V_{\rm XC}({\bf r}) = \frac{\delta E_{\rm XC}[n({\bf r})]}
%      {\delta n({\bf r})}.
%\end{equation}
\end{widetext}

The Schr\"odinger equation (\ref{kohnshameq}) is  solved using the 
Rayleigh-quotient multigrid (RQMG) method \cite{mgarticle1} which
has been implemented in various geometries, 
including the axial symmetry\cite{mgarticle1,mika}.
In the RQMG method, the Rayleigh quotient 
$\left<\psi|H|\psi\right>/\left<\psi|\psi\right>$ 
%$<\psi|H|\psi>/<\psi|\psi>$ 
on the finest level grid is directly minimized, 
the orthogonality constraint being taken into account by a penalty functional.
The Poisson equation (\ref{poisson}) is   solved for the
electrostatic potential $\Phi({\bf r})$ using
a standard multigrid method \cite{brandt1}.

To obtain selfconsistency, we use the simplest possible potential 
mixing scheme,
\begin{equation} \label{simple_mixing}
V_{in}^{i+1} = AV_{out}^i+(1-A)V_{in}^i.
\end{equation}
The largest system of this work contains 2550 electrons, the diameter of
the supercell being $170\;$\AA. 
Obtaining selfconsistency in such a system requires a very small
$A$ value of 0.005. Otherwise the charge sloshing results in divergence. 
More sophisticated mixing strategies\cite{WangPRB01} will be indispensable 
in the future calculations. However, because of the simplicity of our
model systems, we can very accurately estimate an initial guess for 
the selfconsistent effective potential
of large systems using the more easily 
convergent smaller systems as reference. 

In our largest calculation a grid of $319 \times 95$ points is used for
the presentation of the wave functions, potential and density.
Taking into account the unoccupied states needed in the modeling, up to 2400 
different states have to be solved at every selfconsistency iteration.
Luckily, it is straightforward to parallelize the calculation over the 
65 different $m$ values and the two {\bf k}-points (see below). Moreover,
the RQMG method\cite{mgarticle1} 
handles this part of the calculation with optimal  efficiency.

The two-jellium model for the surface
described in Sec. \ref{model}  results in an asymmetric density distribution with a 
surface dipole. Thus the electrostatic potential on the substrate
side is a constant different from that on the vacuum side. 
For the Poisson equation, we thus use  the boundary condition 
of zero derivative on the substrate side, and that of zero value 
on the vacuum side. 
Solving the Poisson equation, the boundaries above and 
below the system are extended to a distance five times greater 
than in the case of the wave functions.

In this work, we calculate the local density of states (LDOS) 
above a surface at distances corresponding to those typical in 
STM measurements (of the order of $20 \;\mathrm{a_0}$). At such distances, 
the amplitude of the wave function decreases by several orders of
magnitude. This kind of modeling is thus a serious test for the
RQMG-method. We have checked the accuracy of our method for % systems
the spherical harmonic oscillator and a model hydrogen atom potential,
for which the wave-functions are known analytically. 
The evanescent tails of  the wave functions solved with the RQMG method agree 
with the analytical ones, even when the amplitude of the wave functions
has dropped by 20 orders of magnitude.
This level of accuracy is beyond the reach of plane-wave methods, where 
periodic boundary conditions are necessary, and which provide
a uniform accuracy across the calculation volume, resulting in spurious 
oscillations in the vacuum parts of the system. 

\section{Modeling the system} \label{model}

We are interested in the system of a monolayer-thick Na QD on the complete
Na monolayer on the Cu(111) surface. We know from experiments that these islands are
approximately hexagonal in shape, following the underlying structure of the 
Na monolayer.

In order to interpret recent STM data for these
types of systems, mapping the energy resolved real-space electron density
near a QD is necessary. 
First, it is of interest to find what level of theoretical modeling is
required. It has been shown previously  \cite{CrampinPRL,CrampinSS,Berndt}
that simple two-dimensional 'particle-in-a-box' calculations give
qualitatively good results, in the sense that the peak structure of LDOS
resembles spectra obtained in the STM $dI/dV$ measurements. In this work we
improve the theoretical description by performing selfconsistent three-dimensional 
DFT calculations,
where the effects of the underlying monolayer and substrate are introduced.
The hexagonal QD is modeled by a cylindrical jellium QD, and the underlying Na
monolayer and Cu(111) substrate by the two-density jellium slab, as described below. Comparisons between calculations of free-standing QD's
and QD's on a substrate show indeed that the underlying monolayer and
substrate induce a new type of states that the simple 'particle-in-a-box'
calculations cannot account for. Since the $z$-dependence of wave-functions
is included in our calculations, we can calculate the tunneling current at realistic STM-tip
distances above the system, and estimate the energy dependence of the step height  from the calculated
constant current topographs.

\begin{figure}
\centerline{\includegraphics[width=8.1cm]{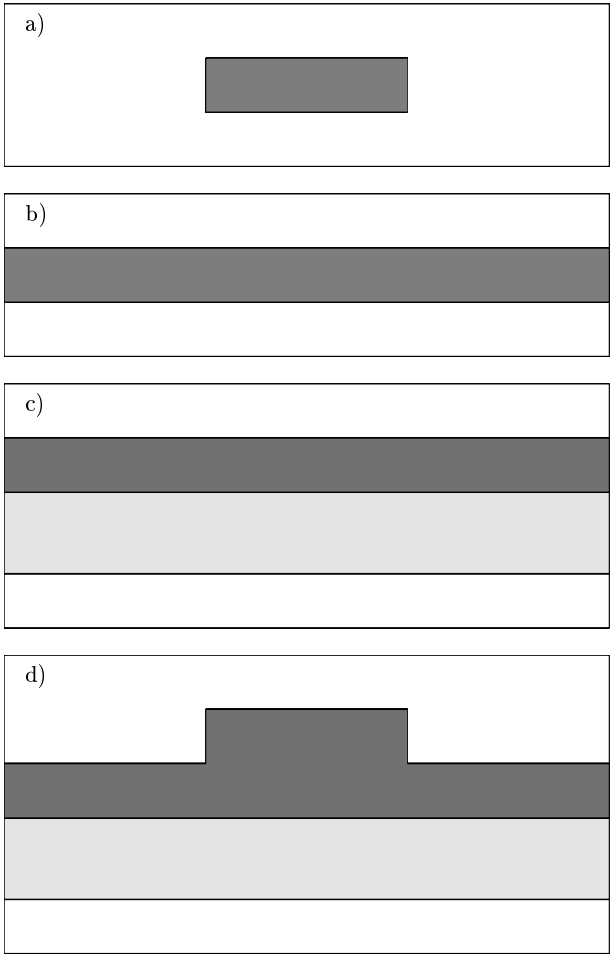}}
%\vspace{0.3cm}
\caption{Profile of the axially symmetric background charge in the case of a) 
jellium model for
free-standing Na quantum dot b) jellium model for free-standing Na monolayer c)
two-jellium model for Na monolayer on Cu(111) d) two-jellium model for Na
quantum dot on Na monolayer on Cu(111)}
\label{Modelfig}
\end{figure}

\begin{figure}
\centerline{\includegraphics[width=8.1cm]{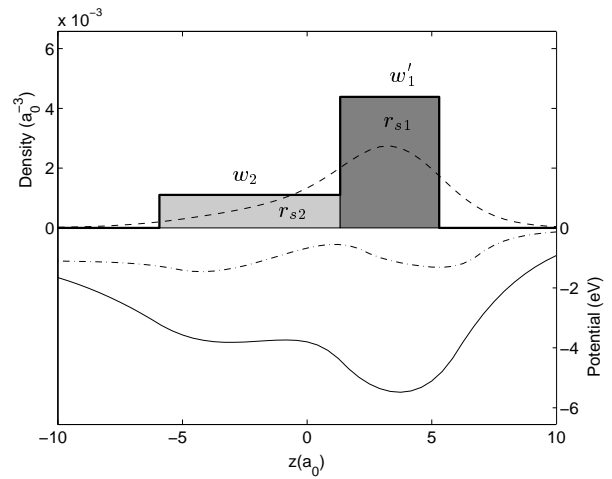}}
%\vspace{0.3cm}
\caption{One complete monolayer of Na on Cu(111) surface within the 
two-jellium model. The positive background charge (shaded areas), electron 
density (dashed line), effective potential (solid line), and electrostatic potential
(dash-dotted line) are shown. The shading corresponds to Fig.~\ref{Modelfig}.
\label{twojelliumfig}}
\end{figure}

The different model systems studied in this work are 
shown in Fig.~\ref{Modelfig}. Our model is readily applicable to 
the case of a free-standing cylindrical quantum dot, where  
we use zero Dirichlet boundary conditions for the wave functions and
for the Coulomb potential. The next step is to model a free-standing 
monolayer. A uniform planar system cannot be exactly reproduced in the 
axial symmetry. We adopt an approximation scheme analogous to the 
Wigner-Seitz method\cite{Wigner-Seitz}. We imagine the plane being filled by 
hexagons, and then approximate these hexagons by area-covering circles.
In order to sample the Brillouin zone of the lattice of circles
we use two ${\bf k}$-points, ${\bf k}=0$ and ${\bf k}$ 
at the Brillouin zone boundary.
The wave functions with ${\bf k}=0$ are required to have a vanishing
radial derivative at the radius of the circle whereas 
the wave functions with ${\bf k}$ at the Brillouin zone boundary
vanish there. According to our calculations the model gives
a  uniform ($r$-independent) charge distribution for the monolayer.
It also minimizes the interactions between a QD inside a 
circle with its periodic images \cite{makov}.
 
The next step is to place the Na monolayer on top of the Cu(111) substrate. 
The effect of the substrate is modeled using the {\it two-jellium model} 
which is illustrated in Fig.~\ref{Modelfig}c  and Fig.~\ref{twojelliumfig}.
We do not model the electrons of the bulk Cu. 
The density of electrons per unit area in the two-jellium model is kept 
the same as in the jellium model for a free-standing monolayer. We add a layer
of lower density jellium, in order to mimic the different wave function decays 
into the substrate and into the vacuum. 
The thickness $w_2$ and density (via $r_{s2}$) give two free parameters 
of the lower density jellium, which we adjust in order to reproduce 
the relevant experimental values of the first  and second 
surface band bottoms 
at the coverages of 1 ML and 2 ML, respectively. Here the first and second 
bands correspond to wave functions with one and two  nodes in the vertical direction, respectively.
 The thickness $w'_1$ of the higher density 
jellium in the two-jellium model is given by
\begin{equation}
w'_1 = w_1 - \left(\frac{r_{s1}}{r_{s2}}\right)^3w_2,
\end{equation}
where $w_1$ is the thickness of the free-standing Na monolayer.

\subsection{The underlying monolayer and substrate}

% Positioned here to fool latex to put where I want it
\begin{figure}
\includegraphics[width=8.1cm]{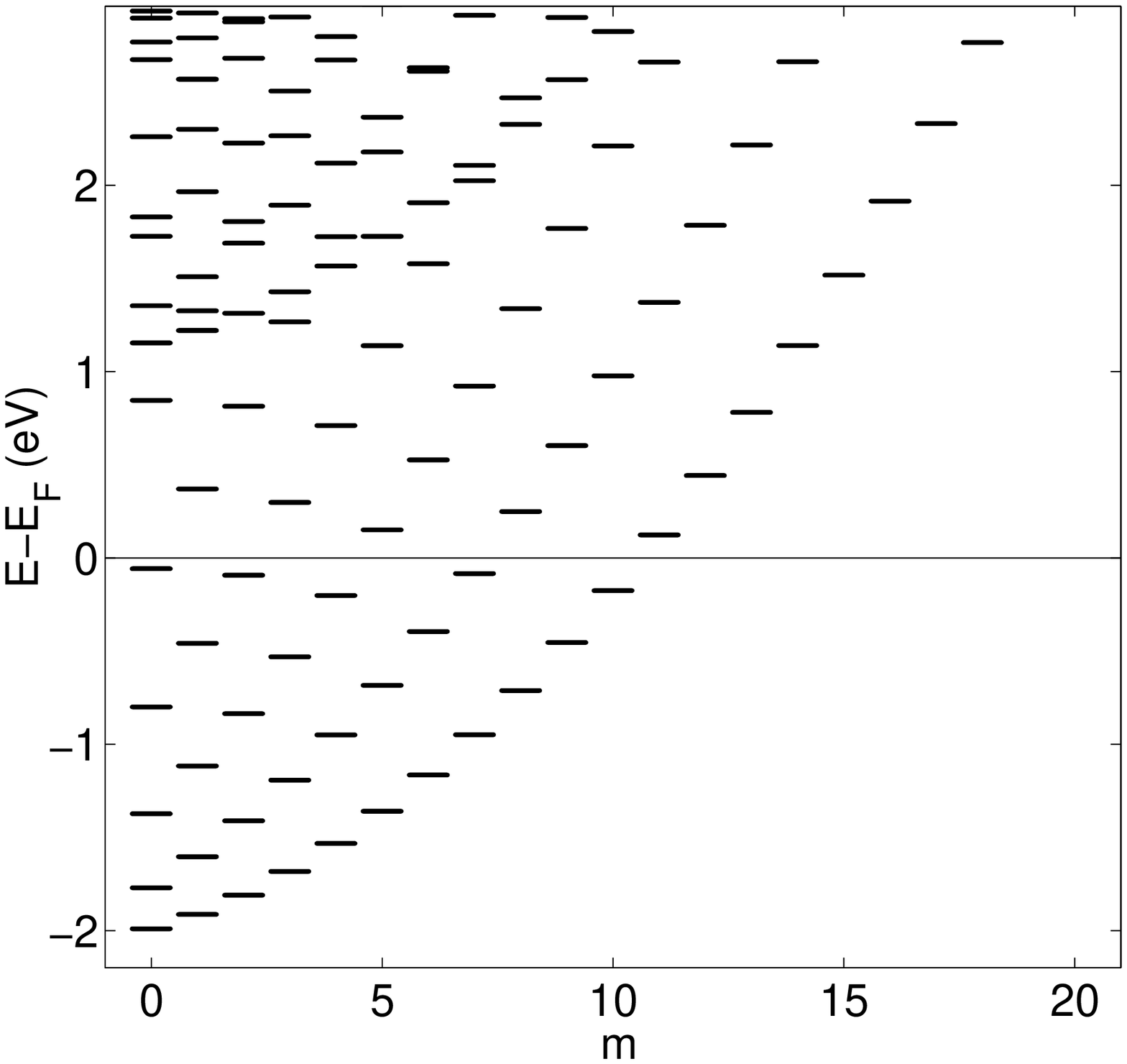}
\includegraphics[width=8.1cm]{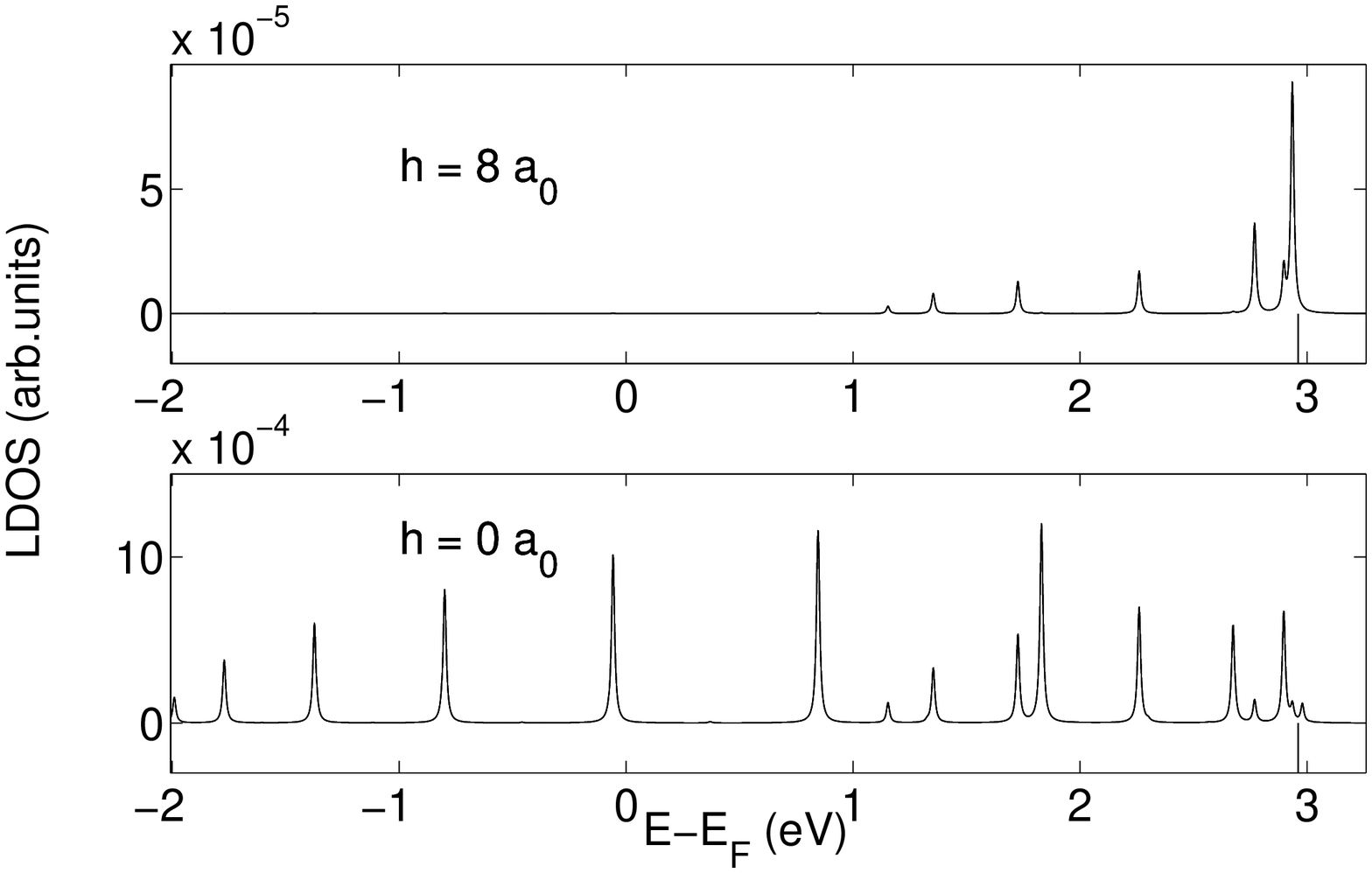}
\caption{\label{en100} The uppermost panel: Energy spectrum of the free-standing 
cylindrical jellium QD (Fig.~\ref{Modelfig}a) containing 100 electrons. 
The discrete eigenenergies are shown as a function of the
quantum number $m$. The top of the figure corresponds to the vacuum level. 
The middle panel:
LDOS calculated at the cylinder axis at $8 \;\mathrm{a_0}$ above the jellium edge.
The lowest panel: 
LDOS calculated at the cylinder axis at the jellium edge. }
\end{figure}

To test our model,  we first study the systems of a free-standing Na monolayer
and that of a  Na monolayer on Cu(111),  and compare the results with other 
theoretical results and experimental findings. The Na jellium density is determined 
from  the bulk nearest neighbor distance of $6.92 \;\mathrm{a_0}$
and the experimental height of $5.5 \;\mathrm{a_0}$
($2.9 \;$\AA) of 1 ML of Na  on  Cu(111)
\cite{Kliewer_Berndt_PRB01,Berndt}. The resulting density parameter, $r_{s}=3.79 \;\mathrm{a_0}$, 
gives a slightly higher density  than its bulk value of $3.93 \;\mathrm{a_0}$ for Na. 
The thickness and the density of the lower density slab have been chosen by 
fitting the bottom of the second band for the 1 ML Na coverage and that of the third
band for the 2 ML Na coverage on Cu(111) to the experimental 
values\cite{Fisher,Lindgren-Wallden88,Carlsson94}.
The values of $r_{s2}=6.0 \;\mathrm{a_0}$ and $w_2=6.3 \;\mathrm{a_0}$ give  
(using the unit cell of radius $72.8 \;\mathrm{a_0}$ containing 400 electrons per monolayer) 
in the 1 ML case the bottom of the second band
at $75\;\mathrm{meV}$  below the Fermi level and in the 2 ML case
the bottom of the third band at $50\;\mathrm{meV}$ above the Fermi level. These values
are reasonably close to the experimental values of about $100\;\mathrm{meV}$ below
and above the Fermi level, respectively \cite{Lindgren-Wallden88,Carlsson94,Carlsson97}. 
The correct positions relative
to the Fermi level are important, because we solve for the electronic structures
selfconsistently, so that the occupancies of the single-electron states 
affect the potential and the character of the states themselves.

\subsection{The quantum dot}
\label{qdot}

\begin{figure}
\includegraphics[width=8.1cm]{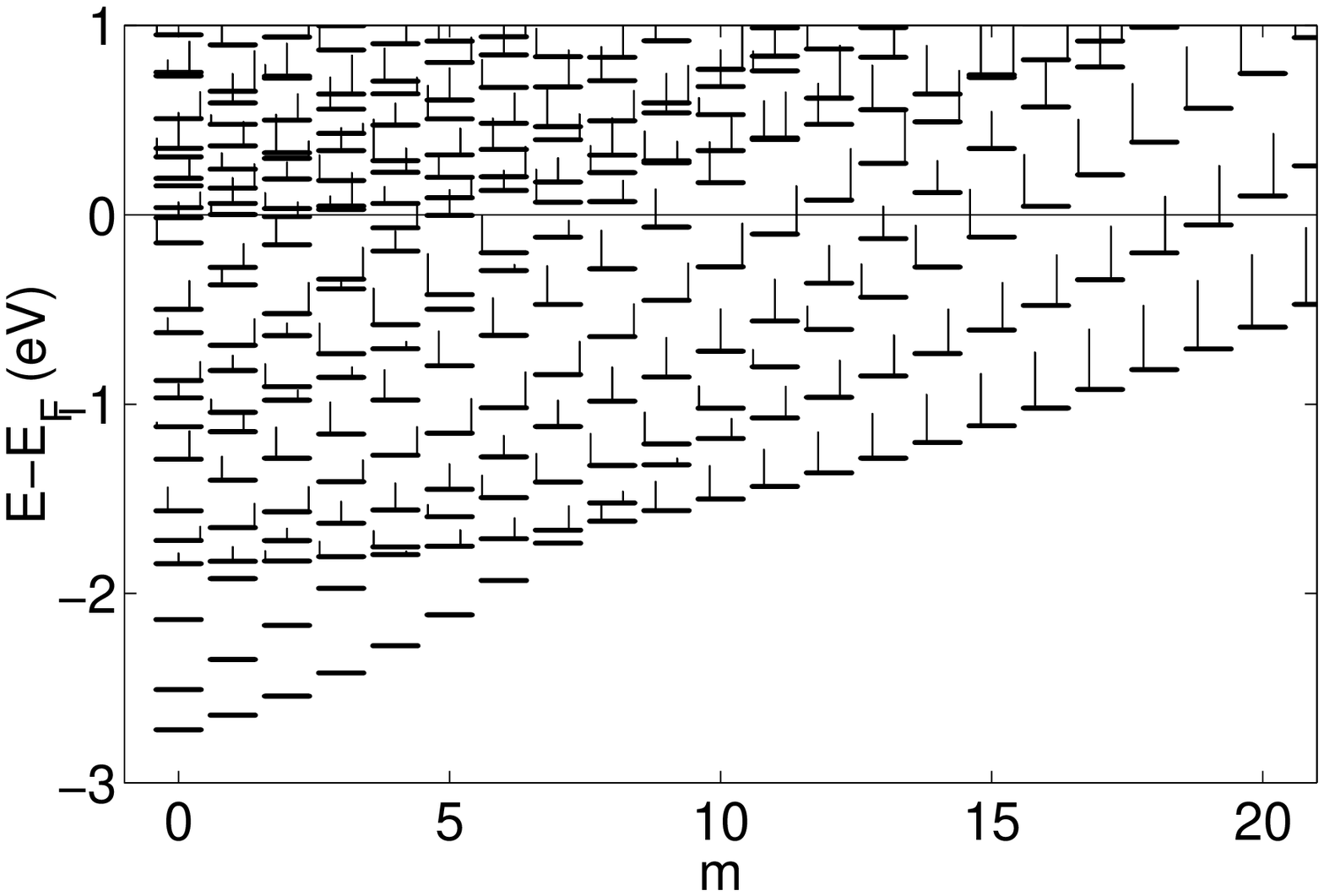}
\includegraphics[width=8.1cm]{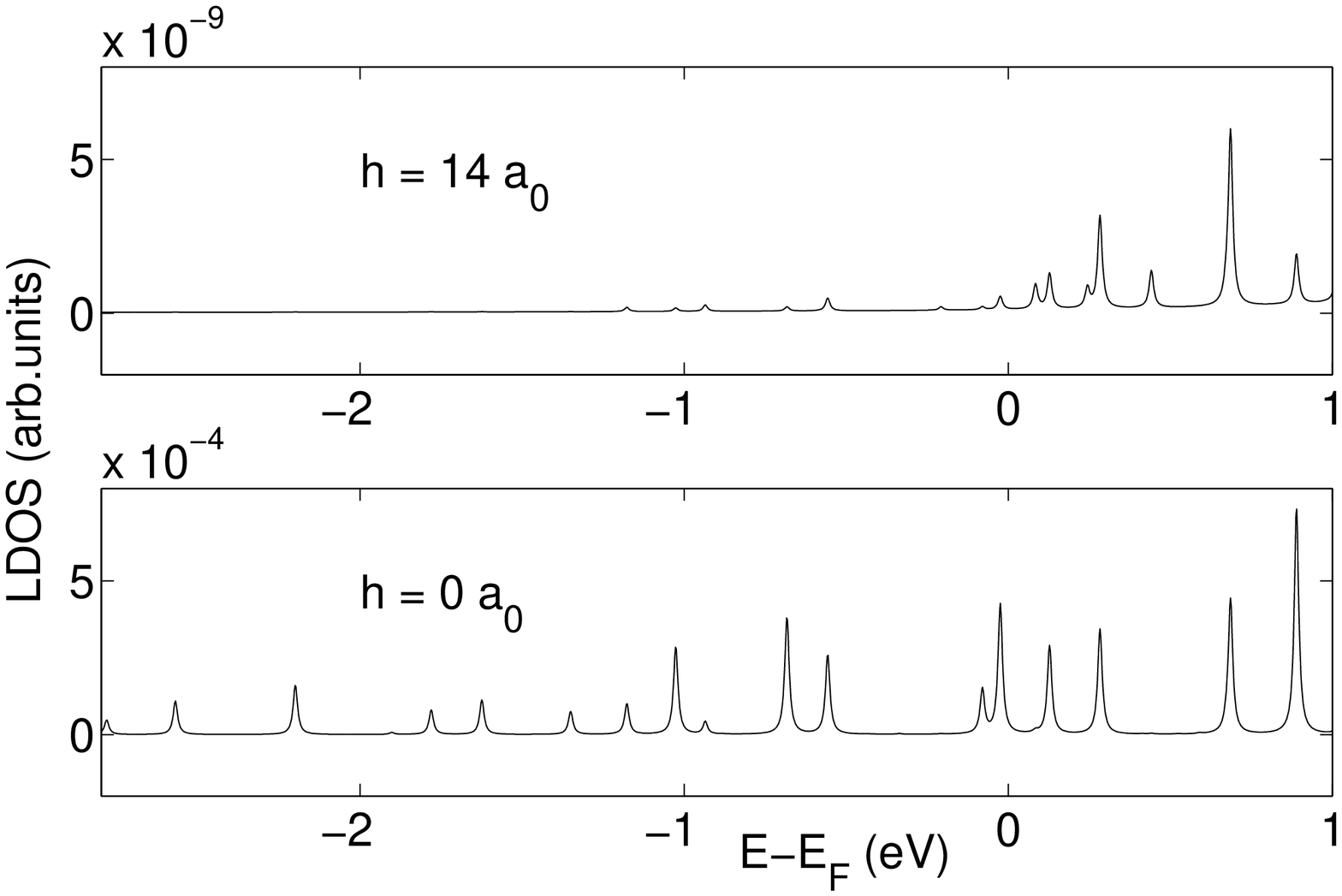}
\caption{\label{en400} The uppermost panel: Energy spectrum of the system of a QD containing 
100 electrons on top of a two-density jellium slab described by a supercell of
400 electrons (Fig.~\ref{Modelfig}d). The ${\bf k}=0$ eigenenergies are given by 
by thick horizontal bars. The thin vertical bars indicate the dispersion in the
${\bf k}$-space. 
The middle panel: 
LDOS calculated at the cylinder axis at $14 \;\mathrm{a_0}$ above the jellium edge.
The lowest panel: 
LDOS calculated at the cylinder axis at the jellium edge. }
\end{figure}

\begin{figure}
\centerline{\includegraphics[width=8.1cm]{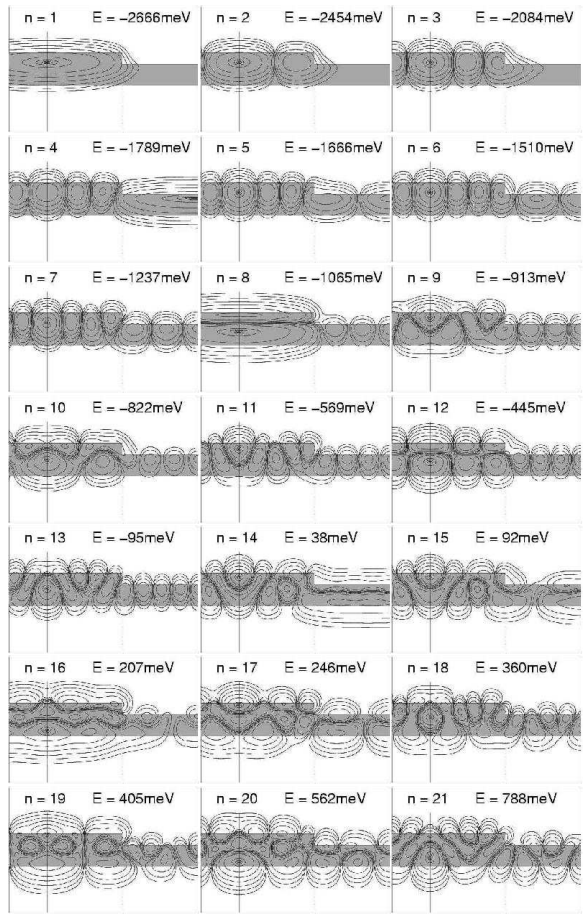}}
\caption{\label{nnn} Wave functions for the 21 lowest lying $m=0$, ${\bf k}=0$ states 
for the system of a monolayer-thick QD, containing 100 electrons, on top of 
a Na/Cu slab (400 electrons per supercell). The wave functions are plotted
in a plane parallel to the z-axis through the center of the QD. In each subfigure,
the cylinder axis is shown by the solid vertical line. The
shading indicates the positive background charge and the dashed vertical 
line points its QD edge. The upper, lower and right-hand subfigureborders 
and the cylinder axis limit the computation volume with the dimension of 
73x60 $\;\mathrm{a_0}^2$. }
\end{figure}

We start by studying a free-standing  monolayer-thick Na QD (Fig.~\ref{Modelfig}a),
since this system shows close resemblance to the simple particle-in-a-box system 
often used as a first approximation when describing the electronic structure of a 
QD on a surface. The number of atoms in the QD is chosen to 100, which corresponds 
to a QD radius of about $R=36.40 \;\mathrm{a_0}$.
The uppermost panel in Fig.~\ref{en100} shows the corresponding energy spectrum
relative to the Fermi energy $E_F$.  Note that the 
the discrete energy eigenvalues are plotted as a function of the quantum number $m$
and not as a function of $\bf k$.	
There are three bands below the vacuum level, 
but only the first band (no horizontal nodal planes) is occupied. The emergence of 
the succeeding second  and third bands can be seen as the condensation of the 
energy levels at around 1.1 eV and 2.5 eV, respectively. 

The LDOS calculated at the cylinder axis at the jellium edge and  at 
$8 \;\mathrm{a_0}$ above the edge are shown in the lowest and the middle panel 
of Fig. 3, respectively. Only states with $m=0$ contribute, since they are
the only ones with nonzero contributions at the axis. The discrete energy levels 
are broadened to Lorenzians with the width $\Gamma = 8\;\mathrm{meV}$.
The LDOS at the jellium edge can easily be resolved in terms of the contributions 
from the different bands: The peaks corresponding to first, second, and third bands
form series with quadratically increasing intervals and smoothly increasing peak 
amplitudes. At the distance of $8 \;\mathrm{a_0}$ above the QD, the
contribution due to the first band states is diminished and the contribution due to the
third band states with high quantization in the z-direction (two horizontal nodal 
planes) is dominating the LDOS. Comparison with the results 
of the simple 'particle-in-a-box' calculation by Lindberg and Hellsing \cite{vanja1} 
shows that these two calculations give qualitatively the same results.

We now compare these results for the free-standing QD to the
system of the QD adsorbed on a Na monolayer on the Cu(111) surface 
(Fig.~\ref{Modelfig}d). In our calculation, the substrate is a 
two-jellium cylindrical supercell containing 400 electrons.
The energy spectrum is shown in the uppermost panel of Fig.~\ref{en400}
in which the different $m$ levels corresponding to the ${\bf k}=0$ -points 
are given with the ${\bf k}$ dispersion calculated using the ${\bf k}$
point at the Brillouin zone boundary. In comparison with the spectrum of the 
monolayer-thick QD in Fig.~\ref{en100} the bands are shifted downwards, because of 
the larger jellium thickness at the QD. The lowest energy states (in the bulb of the 
level diagram) have no ${\bf k}$-dispersion and they are localized at the QD 
and the substrate slab below it (See Fig.~\ref{nnn}). Their dispersion as a function 
of $m$ is similar to that in Fig.~\ref{en100}.
%Introducing the underlying substrate causes new kind of states to emerge which are not 
%localized to the close QD region.
Introducing the underlying substrate gives rise to a new type of states,
which are not localized to the QD region. In fact, these states form 
the overwhelming majority.
 As a result, new bands are induced in the
energy spectrum and they are less dispersive as a function of $m$.
%The reduced $m$ dispersion reflects the fact that the states reside mainly
%outside the QD and the substrate below the QD. 
The reduced $m$ dispersion reflects the fact that the states are extended
over the entire circular supercell. 
For the same reason, these bands 
have a larger dispersion in the ${\bf k}$-space than the localized QD bands.
A simple particle-in-a-box or free-standing 
QD calculation cannot provide states of this kind, and
it is therefore interesting to see to what extent they contribute to the
local electronic structure above the QD.

The LDOS for the QD adsorbed on a Na monolayer on the Cu(111) surface 
is given in the lowest and the middle panel of Fig.~\ref{en400} at the jellium 
edge and  at $14 \;\mathrm{a_0}$ above the edge, respectively. In order
to avoid complications due to the interactions between the supercells we calculate
the LDOS using only the ${\bf k}=0$ states and the LDOS is then calculated
as in the case of the free-standing  monolayer-thick Na QD. To enable a thorough
comparison with the free-standing QD results we have to study first the 
wave functions in more details.

Fig.~\ref{nnn} gives all the wave functions in the interesting energy region
for the $m=0$ states of the QD 
adsorbed on the Na monolayer. The first three states, $n$ = 1, 2, and 3, correspond 
to states within the first band. They are localized to the QD and the substrate slab 
below it and they have no nodes in the z-direction. The $n$ = 4 state shows another
character with a density no longer localized to the QD region but
spread also to the slab region around. This is the first state belonging
to the new type of bands, induced by the slab. State $n$ = 8 is the beginning of the
next band consisting of states with one node in the z-direction (second band in the QD). 
These states are resonance states, the amplitude of which is strongly enhanced in the 
QD region, but due to the hybridization with the delocalized slab states they 
are actually delocalized to the whole system. The $n$ = 14 state starts the next band
consisting of delocalized states with one horizontal node in the slab region,
{\em i.e.} it is a second band in the slab. Finally, state  $n$ = 16 represents the first 
resonance state with two nodes in the z-direction (third band in the QD). In 
Fig.~\ref{nnn} one notes that the states $n$ = 16 and 17 and also the states $n$ = 19 and 20
form pairs. The state lower in energy in the pair is a bonding combination of a
QD state and a surrounding slab state whereas the state higher in energy is an
antibonding combination.

The electronic structure of the QD adsorbed on the Na monolayer, discussed above
in terms of the $m=0$ wave functions, is reflected in the LDOS in Fig.~\ref{en400}. 
First, in the LDOS at the jellium edge we notice that, in comparison with the 
free-standing QD model, the underlying slab both introduces new type of bands and 
squeezes the other bands more tightly together to fit more states below the Fermi 
level (the lowest panels of Figs. 3 and 4). Therefore the present LDOS looks qualitatively 
different from that for the monolayer-thick unsupported QD in Fig.~\ref{en100}.
Moreover, the hybridization of the QD and surrounding slab states to bonding-antibonding 
pairs causes the splitting of peaks seen clearly for the third band states at the distance 
of $14 \;\mathrm{a_0}$ above the jellium edge (the middle panel of Fig. 4). 
If we had a continuous spectrum of slab states we would have a single resonance peak 
with a finite energy width.

Having the modeling of the STM results in mind, the interesting question arising is 
whether or not the localized states calculated by the free-standing QD model give the 
same LDOS  far above the QD as the states in the model including the substrate slab. 
Studying the LDOS plots, matching each peak with the corresponding wave function, we 
notice that the resonance states  with strongly enhanced amplitude in the QD region are 
dominant at large distances above the QD. The contribution of the more delocalized
slab states is small. Therefore the free-standing QD model is expected 
to preserve validity in predicting LDOS at large distances above the QD.
The too broad energy spectrum in the free-standing QD model can be corrected
for by increasing the dot height with a monolayer of Na jellium
or with a two-jellium layer.

\section{Comparison with experiment} 
\label{meas}

\begin{figure}
\centerline{\includegraphics[width=8.1cm]{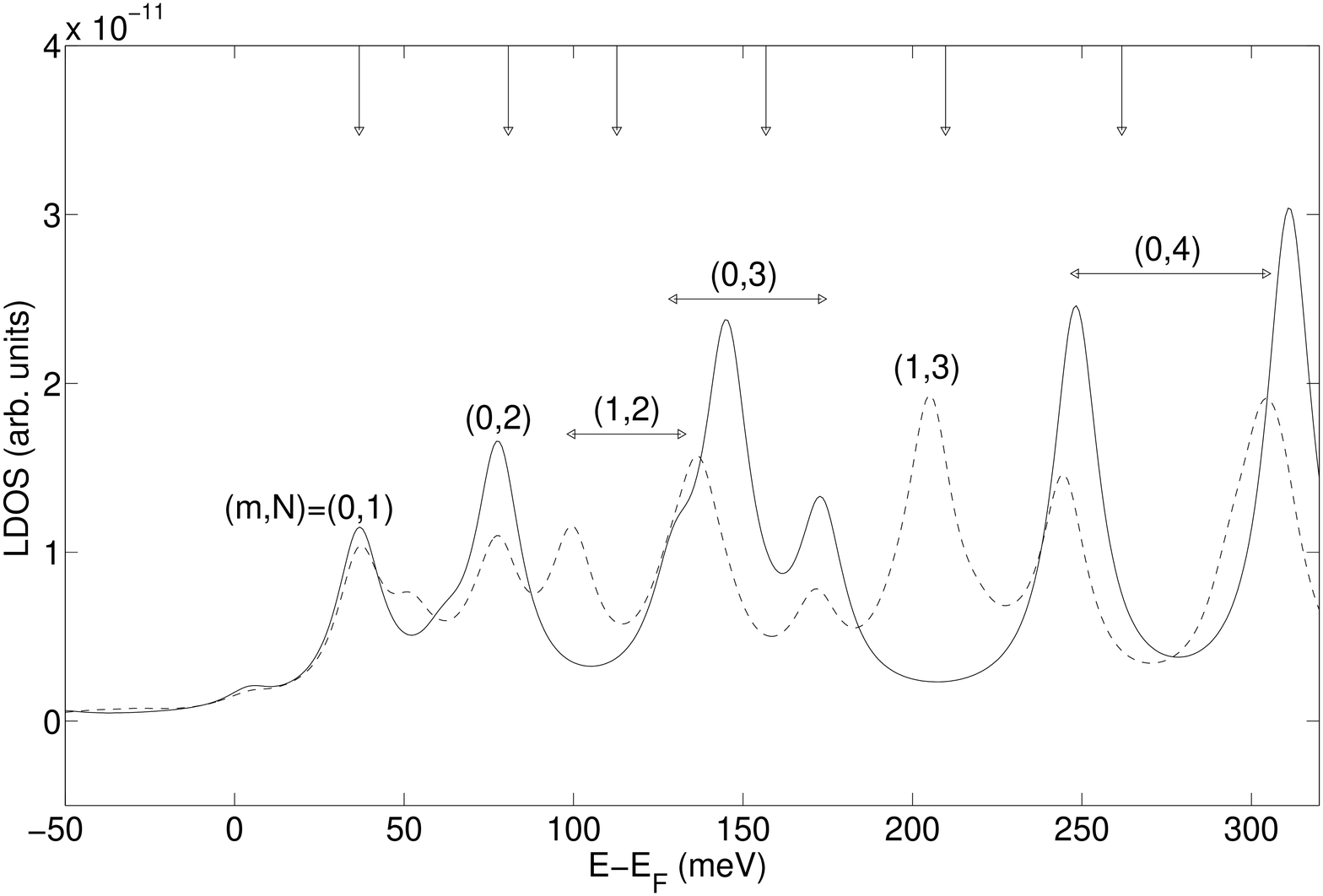}}%\\[.4cm]
\caption{\label{bigg_island_fig_a} Cylindrical QD containing 550 electrons on 
two-jellium substrate. The local density of states is shown at $18 \;\mathrm{a_0}$ above 
jellium edge at the axis  (solid line) and at $r=20 \;\mathrm{a_0}$ (dashed line) away 
from the axis. Lorenz broadening with $\Gamma = 8\;\mathrm{meV}$ has been used.
The relative experimental peak positions\cite{Berndt} are given by vertical arrows 
pointing downwards. The peaks are identified with ($m,N$) resonance states having 
two horizontal node planes in the QD. }
\end{figure}

 \begin{figure}
 \centerline{\includegraphics[width=8.1cm]{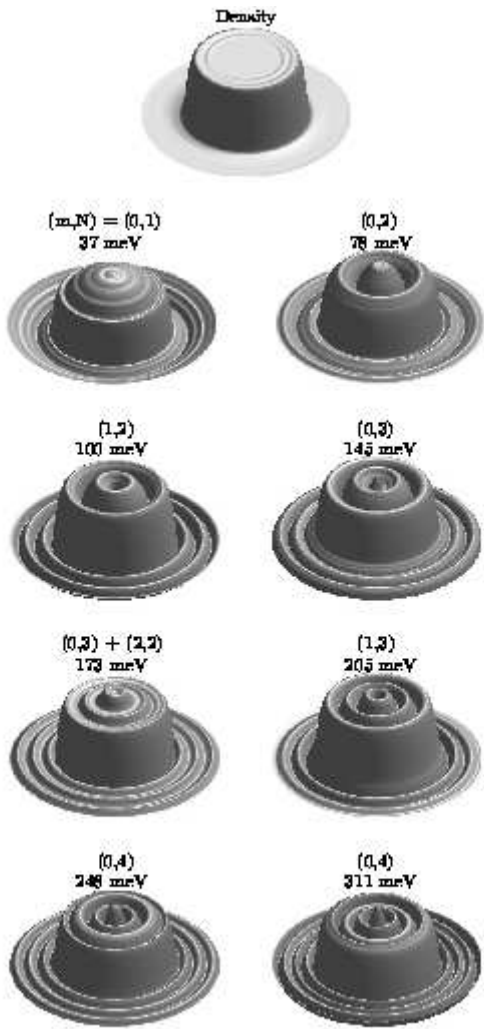}}
\caption{\label{bigg_island_fig_c} Cylindrical QD of 550 electrons on two-jellium substrate.  Isosurfaces of the electron density (top) and the LDOS (with $\Gamma = 0.8\;\mathrm{meV}$) at energies corresponding to the dominant peaks of Fig.~\ref{bigg_island_fig_a}. 
The quantum numbers of the dominant states contributing at each energy is indicated. 
The isovalue for each plot is chosen as the value of the corresponding quantity at $18\;\mathrm{a_0}$ above the jellium edge and$30\;\mathrm{a_0}$ off from the axis. The height-to-radius ratio in the plots is exaggerated. }
 \end{figure}

\begin{figure}
\centerline{\includegraphics[width=8.1cm]{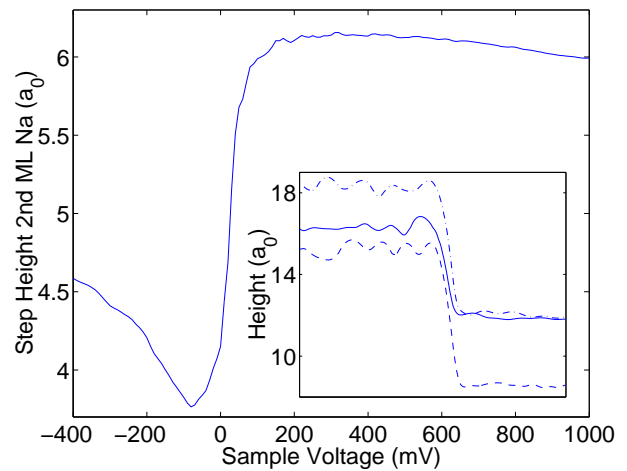}}
\caption{\label{stepfig} Cylindrical QD containing 550 electrons on two-jellium substrate.
The step height of the second Na monolayer determined from calculated constant current surfaces
(see Eq. \ref{current}) is shown as a function of the bias voltage (energy relative to the Fermi level). 
The inset shows the LDOS isosurface profiles (height-to-radius ratio exaggerated) at energies $-400\;\mathrm{meV}$ (dashed line)
$0\;\mathrm{meV}$ (solid line) and $400\;\mathrm{meV}$ (dash-dotted line). The height is measured from the jellium edge of
the second ML QD.
}
\end{figure}

One of the most useful instruments in surface science is the Scanning
Tunneling Microscope (STM) \cite{binnig,tersoff}. It can be used for
measuring the real-space electronic distribution with atomic resolution, as well as the
local energy distribution of electrons and the lifetimes of excited electron
states. The real-space distribution is achieved when scanning the surface
either in the constant current mode, where the tunneling current is kept constant by
changing the tip-surface distance using a feedback mechanism, or in the constant
height mode, where the tunneling current is measured when scanning the surface
at a constant tip height. The resulting image then displays the topography of
the surface.  Information about the local electronic structure is obtained by
measuring the current variation with the applied voltage. This quantity, the
differential conductance $dI/dV$, is proportional to  the product of the 
local density of states (LDOS) and the transmission coefficient $T$\cite{selloni,LiPRB97}. However, if the  applied voltage is small, the bias dependence of
$T$ is small, and with Eq. \ref{densitydef} we approximately have 
\begin{equation}
\label{ldoseq}
\frac{dI}{dV} \propto \sum_{m\bf{k}n} (2-\delta_{0m}) |U_{m\bf{k}n}({\bf r})|^{2}
\delta (\epsilon_{m\bf{k}n}-eV).
\end{equation}
In the STM study by Kliewer and Berndt \cite{Berndt}, constant current topographs
and $dI/dV$ measurements are presented for a Na island on Na monolayer on Cu(111).
The size of the island is $230 \times 170 \;\mathrm{a_0}^2$  
($120 \times 90 \;$\AA$^2$). We have studied a cylindrical
jellium dot with similar dimensions, {\em i.e.} having the radius of $85 \;\mathrm{a_0}$
and containing thus about 550 electrons. The Na/Cu substrate is described in our
calculations by a cylindrical two-density-jellium supercell with the radius of 
$160 \;\mathrm{a_0}$ and containing 2000 electrons. The radius of  $85 \;\mathrm{a_0}$
is actually fixed to reproduce the peak structure of $dI/dV$ spectra by  Kliewer 
and Berndt \cite{Berndt} as well as possible (See Fig.~\ref{bigg_island_fig_a} and discussion below).
For comparison, Kliewer and Berndt \cite{Berndt} used in their modeling two-dimensional
hard-wall hexagons with the radius of $83 \;\mathrm{a_0}$. 

We show in Fig.~\ref{bigg_island_fig_a} the LDOS at $18 \;\mathrm{a_0}$ above the jellium edge
both at the $z$-axis of the  QD and  at $r=20 \;\mathrm{a_0}$ away from the axis. The height 
corresponds to a typical tip-sample distance in the STM experiments. The LDOS is calculated
as for the smaller systems in Sec. \ref{qdot}. The peaks in the figure correspond to 
states with two horizontal nodes in the 2 ML  part of the system (third band in the QD). 
At the axis, only the $m=0$ states contribute, while away from the
axis also peaks with $m \neq 0$ occur. The LDOS peaks can be labeled with the ''quantum number''
$N$ by counting for the number of radial nodes of the corresponding wave functions in the 
2 ML part (See Fig.~\ref{bigg_island_fig_a}). For $m$=0, the $N=1$ state has no 
radial nodes in the dot region whereas $N=2$ has one radial node and so on for larger $N$.
As in the case of the smaller dot discussed above in Sec.~\ref{qdot}, the states strongly peaked in the QD 
are resonance states due to the hybridization with the states of the surrounding monolayer
and span the whole system. Besides the delocalization of the states, the
resonance character causes the fact that in the LDOS (Fig.~\ref{bigg_island_fig_a}) several 
peaks may correspond to the same resonance state as discussed earlier for 
the smaller system. We have identified the LDOS peaks by examining the wave functions.
The horizontal lines below the quantum numbers $m$ and $N$ connect the peaks belonging to the
resonance in question. 

The relative positions of peaks appearing in the experimental $dI/dV$ spectra by
Kliewer and Berndt \cite{Berndt} are shown in Fig.~\ref{bigg_island_fig_a}
as arrows pointing downwards. The experimental data is shifted so that the lowest 
experimental peak coincides with the lowest calculated peak. The experimental spectrum 
is recorded slightly off from the center of the hexagonal QD which should be taken 
into account when comparing with the calculated results. Previous two-dimensional
free-electron calculations for hexagonal potential boxes have reproduced well 
the experimental peak positions \cite{Berndt,CrampinSS}. In the present modeling,
the experimental peak positions agree with the calculated $m=0$ resonance positions
with the exception of the third and fifth experimental peak.
The calculated $m=0$ resonances obey the pattern $\varepsilon_n = E_0 + AN^2$, 
as would be expected for a free particle in a hard-wall cylinder. We have fitted
$E_0\approx 22\;\mathrm{meV}$, $A \approx 15\;\mathrm{meV}$.
It is gratifying to note that in the LDOS recorded off the cylinder axis strong $m=1$
resonance peaks appear so that the third and fifth experimental peaks can be explained.
Thus, our model can reproduce quantitatively the experimental peak positions. According
to our calculations the resonance width increases toward to higher energies. The
increase is maybe slightly stronger than in experiment, indicating a somewhat too weak
confinement of the resonance states in our model.

We have also calculated the isosurfaces of the LDOS at the energies corresponding 
to the dominant peaks in Fig.~\ref{bigg_island_fig_a}.
The results are shown along with the total electron density in 
Fig.~\ref{bigg_island_fig_c}. In order to see clearly the nodal structures of
the different states, the LDOS is calculated using a smaller lorenzian width of 
0.8 meV. The density is smooth in the interior of the QD and shows minor
oscillations at the perimeter of the QD. The development of the nodal 
structure is clear and compares qualitatively well with that found in
the experimental $dI/dV$ maps \cite{Kliewer_Berndt_PRB01}.
It can also be seen that the isosurfaces corresponding to the
two $(m,N) = (0,4)$ peaks differ from each other mainly near the
perimeter of the dot. The higher peak shown at the energy corresponding
to the highest (0,3) state, on the other hand, does not show 
equally clear (0,3)-character. The explanation is a state with
quantum numbers (2,2) at almost exactly the same energy. 
 	
From Eq. \ref{ldoseq} we obtain a simple formula for the tunneling current,
\begin{equation}
I(U,r,z) %= \int_{E_F}^{E_F+U} \frac{dI}{dV}(U,r,z)dE 
\propto  \int_{E_F}^{E_F+U} {\rm LDOS}(E,r,z) dE.
\label{current}
\end{equation}
%From the LDOS isosurfaces,  we can estimate the apparent step height at the
%perimeter of the QD. 
From these numerical constant current topographs, we  can estimate the apparent step height at the
perimeter of the QD. 
The inset of Fig.~\ref{stepfig} shows the constant LDOS
height as a function of the distance from the cylinder axis for the
bias voltages of -400, 0, and +400 meV. 
%The value of the LDOS is the same
%as in the lowest panel of Fig. \ref{bigg_island_fig}. 
The value of the LDOS is $10^{-11}$ arb. units on the scale of
Fig.~\ref{bigg_island_fig_a}. The absolute
height of the isosurface from the jellium edge
depends naturally on the LDOS value chosen, but, according to our 
calculations, the relative
changes are insensitive to the LDOS value over a wide range of values.
In order to construct the apparent step height we first obtain the 
numerical constant current topographs (Eq. \ref{current}), then average the profiles 
over the oscillations above  the 2 ML and 1 ML parts of the system and take 
the difference. The results are shown in Fig.~\ref{stepfig} as a function of the
bias voltage. 
The trends seen can be explained by studying the LDOS-isosurfaces (see the inset of
Fig.~\ref{stepfig}), and then noting that the constant current surfaces
are obtained by simple integation (Eq. \ref{current}). At -100 meV, 
the second band starts to contribute in the 1 ML part of the system rising the height 
there and thereby lowering the step height. Then the onset of the third band
in the 2 ML part rises the step again. This rise is similar to that seen
in the experiment by Kliewer and Berndt\cite{Kliewer_Berndt_PRB01} as well as 
the decline at higher voltages. However, the comparison with the experimental 
result shows differences: our step height is
too low by a factor of 2, and the raising of the step at negative bias voltages
is not seen in the experiment. There may be several reasons for the differences
in the step heights. One is that the experimental step height of 
$5.5 \;\mathrm{a_0}$ ($2.9\;$\AA),
which is determined at a voltage just before the rise in the step height, 
is directly used as the thickness of the jellium describing the second monolayer of Na.
A more consistent procedure might be to take the voltage dependence into account.
Moreover, the apparent step height depends on the relative vacuum decay rates 
of the second  and third band states of the 1 ML and 2 ML systems, respectively.
Their correct description may be too demanding for our simple model.
%The fact that the calculated step height rises at lower negative voltages
%whereas the experimental step height is rather constant at negative voltages
%may reflect the basic difference between the theoretical and experimental
%approaches: The experimental step height profile is determined by keeping
%the tunneling current constant whereas the calculated profile assumes
%a constant LDOS value. Far from the Fermi level these approaches are
%expected to differ \cite{LiPRB97}.

\section{Conclusions} 
\label{conclusion}
In this paper we have presented a model for the electronic
structures of alkali metal islands or quantum dots adsorbed on metal surfaces.
In particular, we have focused on the system of Na on the
Cu(111) surface, where approximately hexagonal Na quantum dots have been 
observed to form during the epitaxial growth of the second Na monolayer.

We have modeled the quantum dots as small cylindrical jellium islands, and the
underlying Na monolayer and Cu substrate as a two-density jellium slab. 
The parameters of the model have been chosen to fit experimental 
spectroscopic data and calculated first-principles band structures for
one and two completed monolayers of Na on the Cu(111) surface.
The calculations were performed in the context of the density-functional
theory, using a real-space electronic structure calculation method.

The calculated results are compared with experimental findings
from scanning tunneling microscope and photoemission experiments.
The model gives local densities of states which are in a quantitative
agreement with constant current topographs and $dI/dV$ spectra
and maps. Thereby the idea of surface states which are localized as
resonances at the quantum dots is supported. The future applications
of the model will include studies of the adsorption and dissociation
of molecules in the vicinity of alkali metal quantum dots.

\acknowledgments

We would like to thank Lars Wallden and S.\AA. Lindgren for sharing
their knowledge on the system based on PES experiments.
We would like to thank R. M. Nieminen for many useful discussions. 
We acknowledge the generous computer resources from the Center for
Scientific Computing, Espoo, Finland. 
One of the authors (T.T.) acknowledges financial support by the Vilho, Yrj\"o and Kalle
V\"ais\"al\"a foundation.  This research has been
supported by the Academy of Finland through its Centers of Excellence
Program (2000-2005).

%\begin{references}

%\begin{figure}
%\centerline{\includegraphics[scale=1.5]{ASA_fig.eps}}
%\vspace{1cm}
%\caption{Modeling planar periodicity in axial symmetry: the honeycomb
%structure of the upper panel is roughly approximated by the circles  of the
%lower panel.}
%\label{asafig}
%\end{figure}


\begin{thebibliography}{99} 
\bibitem{Luth} H. L\"uth, {\em Surfaces and Interfaces of Solids}
(Springer-Verlag, Berlin, 1993).
\bibitem{Gartland} P. O. Gartland, and B. J. Slagsvold, Phys. Rev. B {\bf 12},
4047 (1975).
\bibitem{Heimann} P. Heimann, H. Neddermeyer, and H. F. Roloff, J. Phys. C
{\bf 10}, L17 (1977).
\bibitem{Zangwill} A. Zangwill {\em Physics at Surfaces} (Cambridge
Univ. Press, Cambridge, 1988).
\bibitem{Ashcroft-Mermin} N. W. Ashcroft and N. D. Mermin, {\em Solid State
Physics} (Saunders, Philadelphia, 1976).
\bibitem{Lindgren-Wallden80} S. -\AA. Lindgren and L. Walld\'en, Solid State
Commun. {\bf 34}, 671 (1980)
\bibitem{Lindgren-Wallden87} S. -\AA. Lindgren and L. Walld\'en,
Phys. Rev. Lett. {\bf 59}, 3003 (1987).
\bibitem{Carlsson97} A. Carlsson, B. Hellsing, S.-\AA. Lindgren, and
L. Walld\'en, Phys. Rev. B {\bf 56}, 1593 (1997).
\bibitem{Dudde} R. Dudde, L. S. O. Johansson, and B. Reihl, Phys. Rev. B {\bf
44}, 1198 (1991).
\bibitem{Fisher} N. Fischer, S. Schuppler,  R. Fischer, Th. Fauster, and W. Steinmann,
Phys. Rev. B {\bf 43}, 14722 (1991).
\bibitem{Kliewer00} J. Kliewer, R. Berndt, E.V. Chulkov, V. M. Silkin,
P. M. Echenique, and S. Crampin, Science {\bf 288}, 1399 (2000).
\bibitem{Carlsson} J. M. Carlsson and B. Hellsing, Phys. Rev. B {\bf 61},
13973 (2000).
\bibitem{Hellsing2000} B. Hellsing, J. Carlsson, L. Walld\'en, and
S. -\AA. Lindgren, Phys. Rev. B {\bf 61}, 2343 (2000).
\bibitem{Memmel} N. Memmel and E. Bertel, Phys. Rev. Lett. {\bf 75}, 485
(1995).
\bibitem{Bertel} E. Bertel, P. Roos, and J. Lehmann, Phys. Rev. B, {\bf 52},
R14384 (1995).
\bibitem{Lauritsen} J. V. Lauritsen, S. Helveg, E. L{\ae}gsgaard,
I. Stengaard, B. S. Clausen, H. Tops{\o}, and F. Besenbacher, J. Catal. {\bf
197}, 1 (2001).
\bibitem{Stranick} S. J. Stranick, M. M. Kamna, and P. S. Weiss, Science {\bf
266}, 99 (1994).
\bibitem{Bertel97} E. Bertel, Phys. Status Solidi A {\bf 159}, 235 (1997).
\bibitem{Crommie} M. F. Crommie, C. P. Lutz, and D. M. Eigler, Science {\bf
262}, 218 (1993).
\bibitem{Roder} H. Roder, E. Hahn, H. Brune, J. -P. Bucher, and K. Kern,
Nature {\bf 366}, 141 (1993).
\bibitem{Wimmer} E. Wimmer, J. Phys. F. {\bf 13} 2313 (1983).
\bibitem{vanja1} V. Lindberg and B. Hellsing Surf. Sci, {\bf 506}, 297 (2002).
\bibitem{lang83} N. D. Lang, in {\em Theory of the Inhomogeneous Electron
Gas}, edited by S. Lundqvist and N. H. March (Plenum, New York, 1983) p. 309.
\bibitem{dft} R. O. Jones and O. Gunnarsson, Rev. Mod. Phys.  {\bf 61}, 689
(1989).
\bibitem{mika} T. Torsti, M. Heiskanen, M. J. Puska, and R. M. Nieminen, 
Int. J. Quant. Chem.,  {\em in print}; cond-mat/0205056.
\bibitem{mgarticle1} M. Heiskanen, T. Torsti, M. J. Puska, and R. M. Nieminen, Phys. Rev. B {\bf 63}, 245106 (2001).
\bibitem{WangPRB01} For a short review and a promising new idea,
see the recent paper by D. Raczkowski, A. Canning, and L.W. Wang, Phys. Rev. B, {\bf 64}, 121101 (2001).
\bibitem{Schulte} F. K. Schulte, Surf. Sci. {\bf 55}, 427 (1976).
\bibitem{manninen75} M. Manninen, R. M. Nieminen, P. Hautoj\"arvi, and
J. Arponen, Phys. Rev. B {\bf 12}, 4012 (1975).
\bibitem{heerbrack} W. A. de Heer, Rev. Mod. Phys. {\bf 65}, 611 (1993);
M. Brack, {\em ibid.} {\bf 65}, 677 (1993).
\bibitem{zabala1} N. Zabala, M. J. Puska, and R. M. Nieminen, Phys. Rev. Lett. {\bf 80}, 3336 (1998).
\bibitem{zabala2} N. Zabala, M. J. Puska, and R. M. Nieminen, Phys. Rev. B 59, 12652 (1999).
\bibitem{zabala3} M. J. Puska, E. Ogando, and N. Zabala, Phys. Rev. B 64, 033401 (2001).
%%%
%\bibitem{stab} J. P. Perdew, H. Q. Tran, and E. D. Smith, Phys.  Rev. B {\bf
%42}, 11627 (1990); H.B. Shore and J.H. Rose, Phys. Rev. Lett. {\bf 66}, 2519
%(1991).
%%%
\bibitem{Diehl} R. Diehl and R. McGrath, Surf. Sci. Rep. {\bf 23}, 43 (1996).
\bibitem{Tang} D. Tang, D. McIlroy, X. Shi, C. Su, and D. Heskett,
Surf. Sci. Lett. {\bf 255}, L497, (1991).
\bibitem{Lindgren-Wallden88} S.-\AA. Lindgren and L. Walld\'en, Phys. Rev. B
{\bf 38}, 3060 (1988).
\bibitem{Kliewer_Berndt_SS01} J. Kliewer and R. Berndt, Surf. Sci. {\bf 477}, 250 (2001).
\bibitem{Kliewer_Berndt_PRB01} J. Kliewer and R. Berndt, Phys. Rev. B {\bf 65}, 035412 (2001).
\bibitem{Carlsson94} A. Carlsson, S.-\AA. Lindgren, C. Svensson, and
L. Walld\'en, Phys. Rev. B {\bf 50}, 8926 (1994).
\bibitem{CrampinPRL} J. Li, W.-D. Schneider, R. Berndt, and S. Crampin,
Phys. Rev. Lett. {\bf 80}, 3332 (1998).
\bibitem{CrampinSS} J. Li, W.-D. Schneider, S. Crampin, and R. Berndt,
Surf. Sci. {\bf 422} (1999).
\bibitem{Berndt} J. Kliewer and R. Berndt,
Appl.\ Phys.\ A {\bf 72}, S155 (2001).
\bibitem{brandt1} A. Brandt, Math. Comp. {\bf 31}, 333 (1977).
\bibitem{Wigner-Seitz} C. Kittel, {\em Introduction to Solid State Physics},
7. edition, pp. 248-252.
\bibitem{makov} G.~Makov, R.~Shah, and M.~C.~Payne, Phys. Rev. B {\bf 53}, 15
513 (1996); T.~Korhonen, M.~J.~Puska, and R.~M.~Nieminen, Phys. Rev. B {\bf
54}, 15016 (1996).
%\bibitem{michaelson} H. B. Michaelson, J. Appl. Phys. {\bf 48}, 4729 (1977).
\bibitem{binnig} G. Binnig, and H. Rohrer, Helv. Phys. Acta {\bf 55}, 726 (1982).
\bibitem{tersoff} J. Tersoff, and D. R. Hamann, Phys. Rev. B {\bf 31}, 805 (1985).
\bibitem{selloni} A. Selloni, P. Carnevali, E. Tosatti, and C. D. Chen, Phys. Rev. B. {\bf 31}, 2602 (1985).
\bibitem{xc} D. M. Ceperley and B. J. Alder, Phys. Rev. Lett. {\bf 45}, 566
(1980); J. P. Perdew and A. Zunger, Phys. Rev. B {\bf 23}, 5048 (1981).
\bibitem{LiPRB97} J. Li, W.-D. Schneider, R. Berndt, Phys. Rev. B {\bf 56}, 7656, (1997). 

\end{thebibliography}
\end{document}